# Bibliometric Analysis Of Herding Behavior In Times Of Crisis

Fenny Marietza[1], Ridwan Nurazi[2], Fitri Santi[3], Saiful[4]

Economic and Business Faculty, University of Bengkulu,
Email: Mari3tza@gmail.com

*Abstract*

*The social and psychological concept of herding behavior provides a suitable solution to give an understanding of the behavioral biases that often occur in the capital market, especially the herding behavior of investors in times of crisis. The aim of this paper is to provide an overview of the broader bibliometric literature on the term and concept of "herding behavior". Articles are collected through the help of software consisting of Publish or Perish (PoP), Google Scholar, Mendeley and VOSViewer through a systematic approach, explicit and reproductive methods. In addition, the articles were scanned by Scimagojr.com (Q1, Q2, Q3 and Q4), analyzing 83 articles of 261 related articles from reputable and non-reputable journals from 1996 to 2021.*

*Mendeley software is used to manage and resume references. To review this database, a classification was performed using the VOSviewer software. Four clusters were reviewed; The words that appear most often in each group are the type of stock market, the type of crisis and the factors that cause herding. Thus these four clusters became the main research themes on the topic of herding in times of crisis. Meanwhile, methodology and strategy are the themes for future research in the future.*

## Introduction

Some research that discusses the herding behavior of investors has discussed that herding behavior occurs in some markets. Economou, Kostakis and Philippas (2011) found during the crisis of 1998 to 2008 in Greece there had been herding behavior, as well as the stock market in Italy. Similarly, Dang and Lin's research (2016) obtained evidence that herding behavior occurs in the capital market, where herding behavior will tend to increase in the period of increase in the share price compared to the decrease in the share price. Espinosa, Crishtian and Arias research also found the presence of vulture behaviour in the Australian stock market. Li, Hu and Tang (2019) conducted research on Chinese stocks during three periods of crisis and found that herding behavior in China occurred for type A stocks only. Vulture behavior in Europe caused by the financial crisis in 2007-2008 has been proven to occur (Ouarda, El Bouri and Bernard, 2013). In addition, some researchers also say that herding behavior occurs in the stock markets of Japan, Indonesia, Hongkong, Vietnam, and the Balkans(Nakagawa, Oiwa and Takeda, 2012; Bui, Nguyen and Nguyen, 2015; Teng and Liu, 2014; Vo and Phan, 2017, Economou, 2019). From some of these studies it can be said that herding behavior can be found in advanced stock markets, developing and borders.

Herding behavior occurs a lot in countries that are experiencing pressures and crises. Like the global financial crisis in 1997 -1998, the financial crisis of 2007-2008, 2015. When the stock market moves strongly and investors will tend to ignore personal confidence and follow other investors. Herding behavior will occur when stock market movements are sent with a positive or negative return. (Dang and Lin, 2016; Economou, Kostakis and Philippas, 2011; Emekter, 2020; Blasco, Navidet et.al, 2017; Huang, Lin and Yang, 2015; and Ouarda, El Bouri and Bernard, 2013). During Europe's crisis, this herding behavior was also discovered by some peeliti. Some countries in Europe are proven to experience dispersion returns on its stock market and cause herding behavior. Research discussing herding behavior during the crisis of disease outbreaks was also found by several researchers such as, Chang, Mc Aleer and Wang (2020) and Espinosa, Christian and Arias (2020) and Luu and Luong (2020) who found that during the covid 19 pandemic that caused pressure on the market, causing herding behavior in the stock market. Based on some previous research, it can be concluded that herding behavior occurred during the global financial crisis, financial crisis in certain regions and pandemic disease outbreaks that can affect investor psychology in making investment decisions.

Previous research also presents some facts that there are several things that are the determining factors of the occurrence of herding behavior. Those factors include (i) extreme stock returns (Bohl, Klein and Siklos, 2014; Dang and Lin, 2016; Economou, Karsikas, and Vickers, 2016; Bharti, et.al, 2020), (ii) investor sentiment (Zouaoui, 2011; Philippas,Economou, Babalos and Kostakis, 2013; Abbes, 2013; Teng and Liu, 2014; Mokni, 2020), (iii) market sentiment (Blascoa, Curredorb, and Ferreruela, 2011), (iv) financial analyst recommendations (Lin, 2018), (v) forecasters (Pierdzioch, Rulke, and Stadtmann, 2012; Tsuchiya, 2020), (vi) presence of foreign investors ( Park (1995) and Sachs (1998) Frankel and Schmukler (1996, 1998) Bowe, Domuta (2004); Chen, Yang and Lin (2012), (vii) institutional investors (Nakagawa, Oiwa and Takeda (2012), (viii) cascade information (Chari and Kehoe (2004), (ix) macro and micro news released by the government (Galariotis, Rong and Spyrou (2015); Hwang and Salmon (2004); Luo and Schinckus (2015); Mesis and Zapranis (2014), (x) fears felt by investors (Huang and Wang (2017), (xi) market pressures (Stavroyiannis and Babalos (2017) Demirer and Zhang (2018), Stavroyiannis and Babalos (2019) Junior and Palazzi (2020).

So far there has been no bibliometric analysis that discusses herding behavior that occurs in times of crisis. The purpose of this research is to provide a broader literature review of the terms and concepts of herding behavior in times of crisis, in order to answer the following questions:
1. How are articles of herding behavior in times of crisis classified?
2. What is the growing trend of research on herding behavior for this time of crisis? Which topics are more widely published subjects?
3. What are the topics regarding herding in times of crisis for future research that provide opportunities for more research?

This paper is prepared as follows: (1) provides an overview of the literature on herding behavior based on previous research in part 1. In part 2, the definition of herding behavior and the review of the term herding behavior exist. The methodology used to perform bibliometric analysis, including method steps related to the use of publish or perish (PoP) software will be presented in part 3. Part 4 will present the results using VOSviewer. The recommendations, conclusions and limitations of the study will be explained in section 5.

**Literature Review**
**Definition of Herding Behavior**

Some researchers propose different definitions of herding behavior. Vives (2008) said that herding behavior captures the suitability of investor choice. Institutional investors will imitate each other when making an investment decision. Christie and Huang (1995) define vulture behavior as the behavior of investors who do not make investment decisions based on their rational analysis but follow the actions of others. Thus, it can be said that herding behavior is a consequence of social pressure and the general logic that group voices are better than individual voices. Group behavior will be able to destabilize the situation in the market, by pushing the share price away from its fundamental value and causing market inefficiency.

Christie and Huang (1995) said that vulture behavior is an environment where investors follow the group's decisions even if it is different from the information it has itself. This herding behavior by investors can destabilize the market and investors will act based on judgments made by others. Another definition was expressed by Bikchandani and Sharma (2001) which defines herding behavior as rational as a form of behavior imitating other investors that can cause stock price volatility. Furthermore, Bikchandani and Sharma say that herding behavior is also false or (spurious) which refers to the act of following a group of investors who have important information and earn maximum profit. There are three reasons why investors engage in herding behavior, namely incomplete information, investor reputation and compensation.

Lakonishok et al (1992) defines vultures as the average tendency of money management groups to buy or sell shares. Nofsinger and Sias (1999) say that vulture behavior is a collective behavior that arises from uncoordinated individual choices. Herding behavior will refer to a group of investors who trade in the same direction over several periods.

From some of the above definitions it can be said that there are two main streams of empirical studies that investigate vulture behavior in financial markets. Vultures can be detected from market returns and vultures are focused on institutional investor behavior.

**Table 1. Defining Herding Behavior**

| Researchers | Mimic Behavior | Rational Action | Irrational Action | Institutional Investors | Stock Price Volatility |
|---|---|---|---|---|---|
| Vives (2008) | V | | | V | |
| Christie &Huang (1995) | | | V | | V |
| Bikchandani & Sharma (2001) | V | V | V | | V |
| Lakonishok et al.(1992) | | V | | | |
| Nofsinger & Sias (1999) | V | V | | | |

Source: Previous research data

**Research Method**

A library review should be conducted using systematic, explyptic and reproductive methods (Fink, 2005; Garza-Reyes, 2015), or thought mapping method that emphasizes the limits of knowledge (Tranfield et al., 2003). This bibliometric review is used in disciplines that cover quantitative studies in the form of papers, articles, journals, books or other types of written communication (Heersmink et al., 2010). In this study, bibliometric analysis was conducted with five steps introduced by Fahimnia et al. (2015). The steps of this study are to search for keywords about vulture behavior in times of crisis, initial search results, improvement of search results, preparation of initial data statistics and data analysis that will be described in the following subsections.

1. Defining search keywords

   This literature search on herding behavior was conducted in September 2020, using the keyword "vulture behavior in times of crisis". PoP software is used to collect this data. At the beginning of the search, it is done by entering keywords into pop software that is specialized in scopus indexed journal searches only, and setting special conditions for "journal/article", "title only words" and 0-0 years. In search, for newspapers, book reviews and book chapters are excluded in the search. From this database obtained 165 articles from 1996 to 2021.

   Initial search results

   During the initial search, no year span was set, so it was obtained that the oldest journals were discussing the behavior of vultures in times of crisis. So it was obtained that research that discusses the behavior of vultures in times of crisis there began in 1996. The following is data based on identification of PoP.

**Table 2. Top Ten Articles Identified From POP (search unfiltered)**

| Researchers | Title | Year |
|---|---|---|
| S Bikhchandani, S Sharma G. Kaminsky | On crises, contagion, and confusion | 2000 |
| T. Chiang | An empirical analysis of herd behavior in global stock markets | 2010 |
| S. Hwang | Market stress and herding | 2004 |
| G. Kaminsky | What triggers market jitters?: A chronicle of the Asian crisis | 1999 |
| G. Calvo | Mexico's balance-of-payments crisis: A chronicle of a death foretold | 1996 |
| W. Kim | Foreign portfolio investors before and during a crisis | 2002 |
| F. Economou | Cross-country effects in herding behaviour: Evidence from four south European | 2006 |
| E.C. Galariotis | Herding on fundamental information: A comparative study | 2015 |
| D. Kenourgios | Equity market integration in emerging Balkan markets | 2011 |
| N. Philippas | Herding behavior in REITs: Novel tests and the role of financial crisis | 2013 |

Source: PoP Data

2. Improved search results

The next step is to issue an article that is not suitable for screening criteria. The table below shows the results for references to articles that seem important to meet the requirements.

**Table 3 Detailed Search Filtering Criteria**

| Search filtered | Number of Articles |
|---|---|
| Irrelevant | 43 articles |
| Non-English Languages | 2 articles |
| Review /Book/Conference | 23 articles |
| Total | 68 articles |

Source: PoP software

**Table 4. Comparison Metrics**

| Metrics Data | Initial Search | Narrow Search |
|---|---|---|
| **Query** | Journals, articles, erratum, reviews and conference, herding behavior in times of crisis | Journals, articles, herding behavior in times of crisis |
| **Source** | Scopus | Scopus |
| **Paper** | 165 | 98 |
| **Citations** | 3.242 | 2.422 |
| **Years** | 1996-2021 | 1996-2021 |
| **Citations_Years** | 130,16 | 96,88 |
| **Citation_articels** | 19,84 | 24,71 |

| Researchers_articles | 0,99 | 1 |
|---|---|---|
| h_index | 25 | 22 |
| g_index | 54 | 48 |
| hA_index | 9 | 9 |
| hI_annual | 1 | 0,88 |
| hI_norm | 25 | 22 |

Source: PoP software

3. Build initial statistics

Searches are generated after repair, stored in Mendeley software, and exported to RIS format to include all important information related to the article, including title, author name, abstract, keywords and journal specifications (journal publications, speeding years, volumes, publications and pages). The data is analyzed so that articles can be classified by year and source of publication and publisher. From table 3 it is known that the articles published are from the journal scopus Q1 through Q4. The figure below shows the number of articles published from year to year. Figure 1 describes journals from years (1996-2021) and distribution of publications per year.

Regarding the publishing group, Elsevier appears most often followed by Emerald insight, Taylor and Francis, and Routledge (Figure 2).

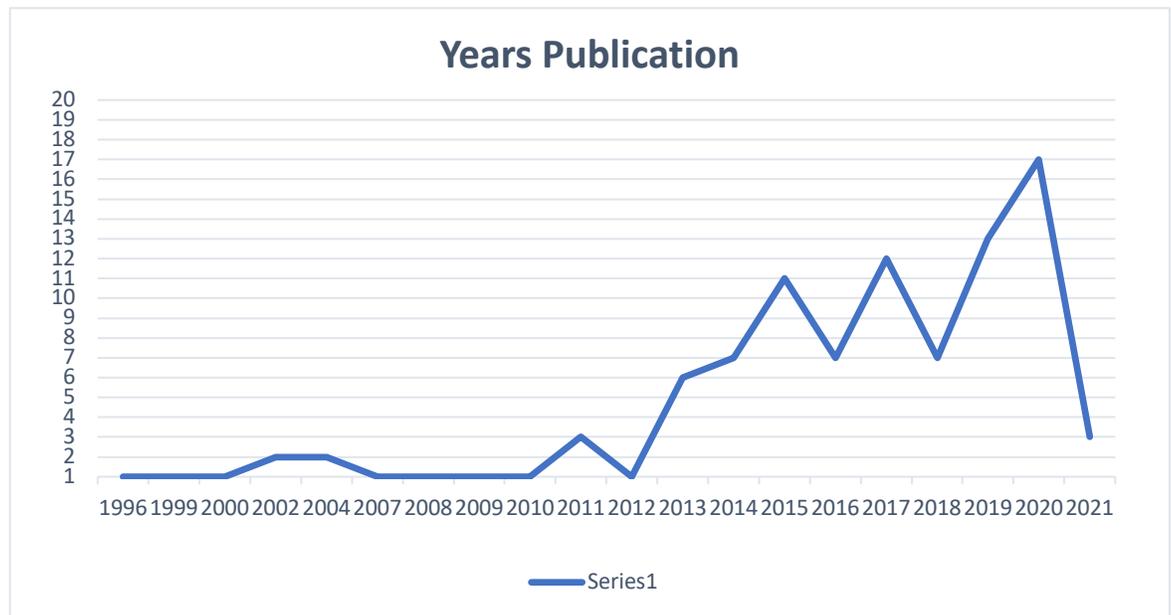

**Figure 1 Years of Publication**

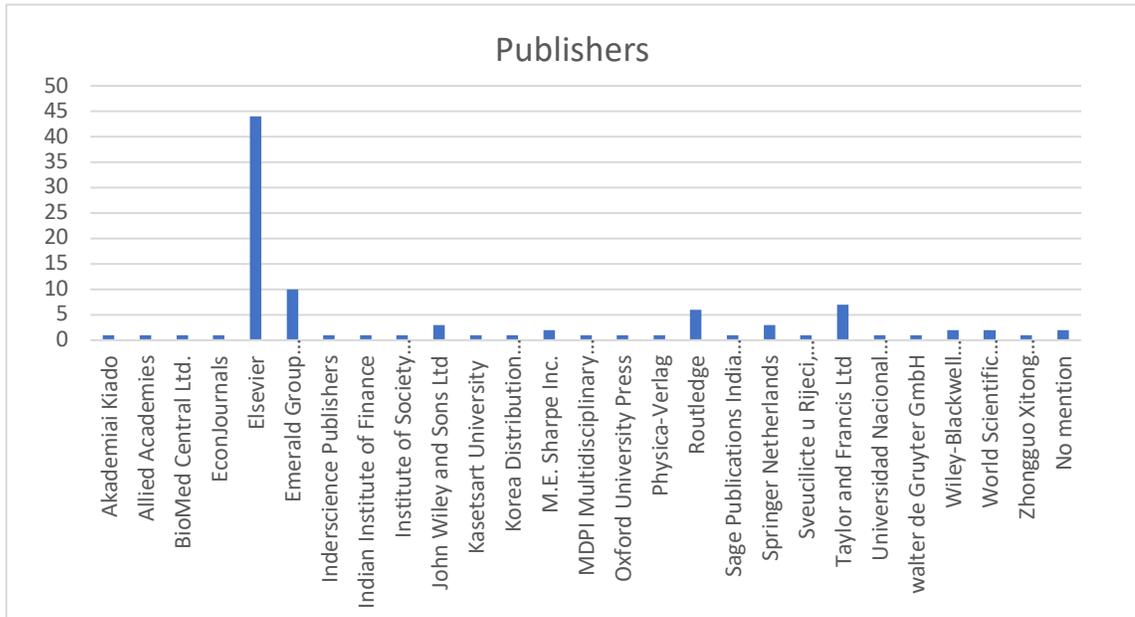

**Figure 2 Article Publisher Mapping**

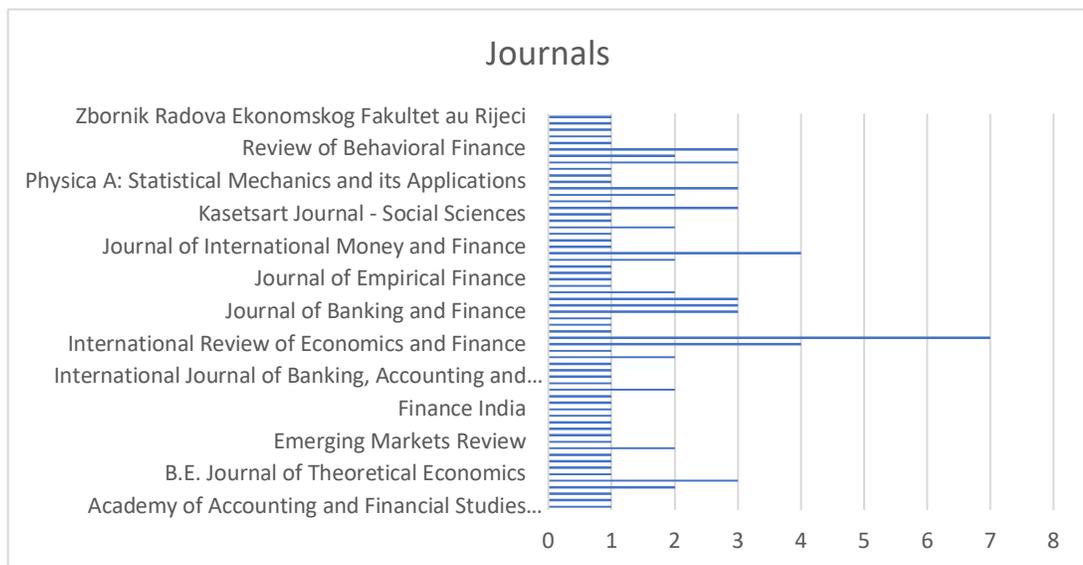

**Figure 3 Journal Mapping**

A total of 61 journals to date published one article themed on herding behavior in times of crisis, 61 of them are ' *Academy of Accounting and Financial Studies Journal', 'Accounting and Finance', 'Acta Oeconomica', 'Applied Economics', 'Applied Economics Letters',' B.E. Journal of Theoretical Economics', 'Contaduria y Administracion', 'Economic Annals-XXI', 'Economic Modelling', ' Emerging*

*Market Finance and Trade', Emerging Market Review', Empirical economics, European Financial Management', 'European Journal of Finance', 'Finance India', 'Finance Research Letters', Financial Innovation', 'Global Business Review', 'International Journal for Equity in Health', 'International Journal of Banking, Accounting and Finance', 'International Journal of Economics and Financial Issues', ' International Journal of Emerging Markets', 'International Journal of Finance and Economics', 'International Journal of Modern Physics C', 'International Review of Economics and Finance', 'International Review of Financial Analysis', 'Journal of Applied Business Research', 'Journal of Asian Economics', 'Journal of Asian Finance, Economics and Business', 'Journal of Banking and Finance', 'Journal of Behavioral and Experimental Finance', 'Journal of Behavioral Finance', 'Journal of Business Research', 'Journal of Economic Studies', 'Journal of Empirical Finance', 'Journal of Financial Economics', 'Journal of Interdisciplinary Mathematics', 'Journal of International Economics', 'Journal of International Financial Markets, Institutions and Money', 'Journal of International Money and Finance', 'Journal of Multinational Financial Management', 'Journal of Open Innovation: Technology, Market, and Complexity', 'Journal of Real Estate Finance and Economics', 'Journal of Risk Finance', 'Kasetsart Journal - Social Sciences', 'Managerial Finance', ' Maritime Policy and Management', 'North American Journal of Economics and Finance', 'Pacific Basin Finance Journal', 'Physica A: Statistical Mechanics and its Applications', 'Quantitative Finance', 'Renewable and Sustainable Energy Reviews', 'Research in International Business and Finance', 'Review of Accounting and Finance', 'Review of Behavioral Finance', 'Review of Pacific Basin Financial Markets and Policies', 'Small Business Economics', 'Studies in Economics and Finance', 'Xitong Gongcheng Lilun yu Shijian/System Engineering Theory and Practice', dan 'Zbornik Radova Ekonomskog Fakultet au Rijeci'*

4. Data Analysis

   This study presents a bibliometric problem for the term herding behavior in times of crisis from a database. The bibliometric review in this study used PoP software which was developed and launched in 2006 (Harzing, 2011). For this analysis, PoP version 5.28.1.6296 was used. This software was used on September 1, 2020, obtained 165 initial articles and refined the results leaving only 98 articles. These findings suggest that journals with Scopus ratings of Q1 and Q2 have the most significant impact on citations compared to other journals.

## Results And Discussion

This research shows that the journal Q1/Q2 has a substantial impact on the metrics associated with citations. Articles written by Chiang (2010) and Hwang (2004) have been widely cited by 309 and 238 other authors, and become the most frequently cited articles in the study. The article is titled 'An empirical analysis of herd behavior in global stock markets' and 'Market stress and herding'. For data on all articles cited more than 50 times (with complete information) is in table 5.

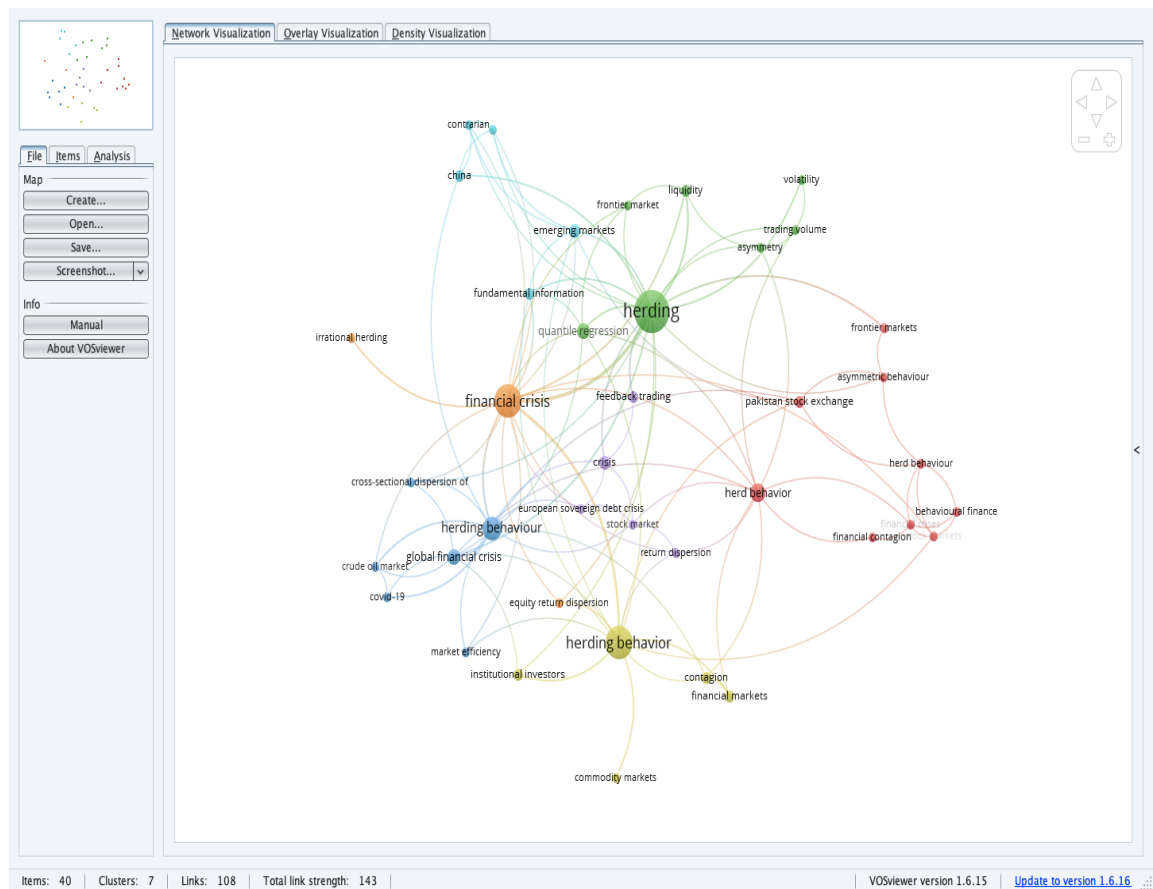

**Figure 4 Mapping of research network visualization**

After taking into account the frequency of citations and other metrics, the study also analyzed the output of PoP software into VOSviewer software to determine what keywords appeared most often. Keyword frequencies can be set by 1, 5, 10, 20 or other events. The VOSviewer tool was developed by Van Eck and Waltman in 2010 (see http://www.vosviewer.com) and can be used to visualize bibliometric maps. This device shows bibliometric mapping on three different visualizations namely network visualization, overlay visualization, and density visualization. VOSviewer can classify

keywords into different clusters. Each cluster shows the weight of its occurrence or the appearance of the term and this explanation answers the first research question.

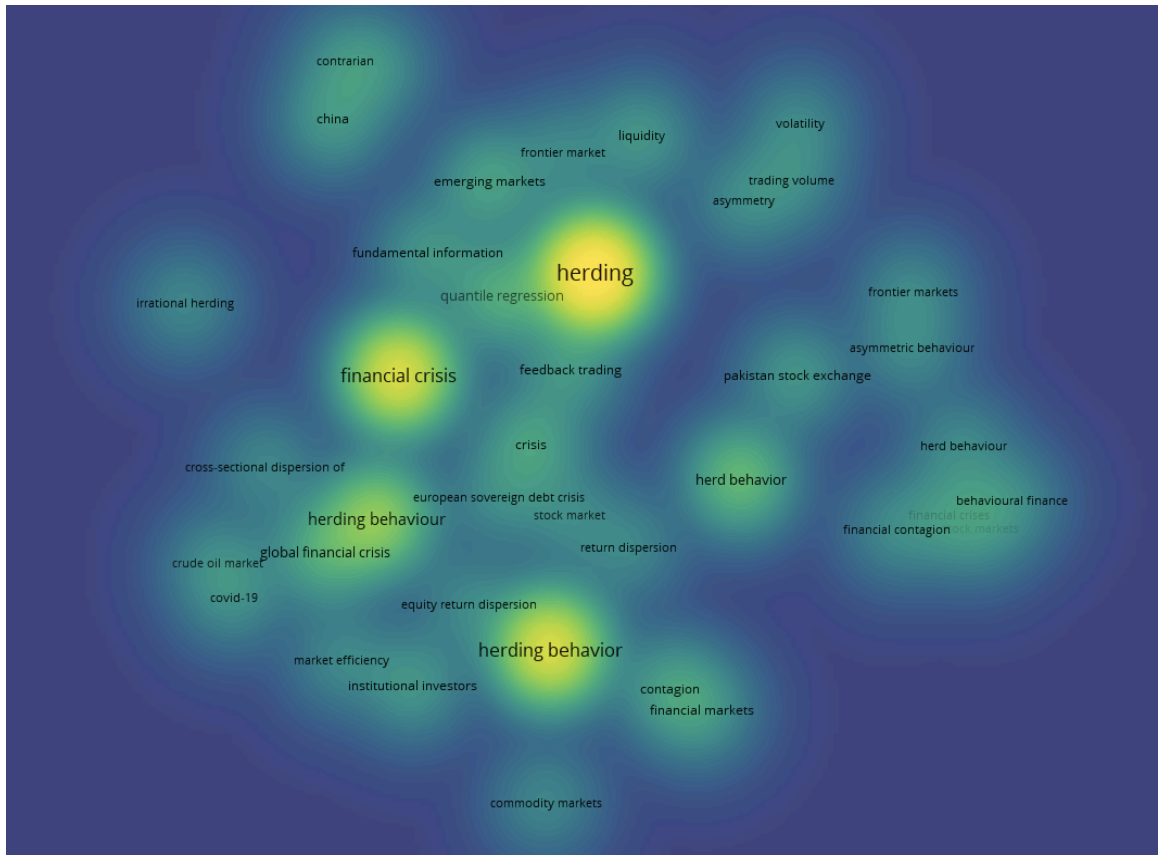

**Figure 5 Mapping the research density visualization**

Extracting from existing titles and abstract fields, we get 240 terms and 40 items. Seven clusters are identified here. The first cluster in figure 4 consists of 9 items with the most common term being herd (9 repetitions). The second cluster has 7 items with the most common terms appearing are vultures (35 times), liquidity (3 times) and quantil regression (5 times). The third group identified 6 items with the most emerging terms being cross sectional disperse (4 times), global financial crisis (5 times), herding behavior (11 times). The fourth cluster consists of 5 items with the most emerging terms are contagion (3 times), financial markets (3 times), herding behavior (21 times), institutional investors (3 times). For cluster five consisting of 5 items with the most frequent terms are Crisis (4 times), European sovereign debt (3 times), feedback trading (3 times), cluster six consists of 5 items with the most frequent terms / appear is the stock market developing (4 times), basic information (3 times). The last is a cluster of seven with 3 items in it, and the most common term is financial crisis (21 times).

Keywords in each cluster represent a stream of herding behavior research in times of crisis. More detailed information is presented in table 6. Each cluster shows a trend of studies related to herding behavior in times of crisis that can be represented through the emergence of these terms. This data makes it possible to answer the second research question, "what is the trend of herding behavior research in times of crisis". 'herd', 'vulture', 'liquidity', 'cross sectional disperse', 'financial crisis', 'vulture contagion', 'financial markets', 'institutional investors', 'feedback trading', 'emerging stock markets are the most common words; 'asymmetry behavior', 'trade volume', 'Covid 19 pandemic', 'Contrarian', 'momentum' are some of the words that are rarely used and can be researched in future research. A variety of topics can be developed based on these keywords.

**Table 5 Articles with more than 50 citations**

| No | Citations | Per Year | Author | Title | Year | Publication |
|---|---|---|---|---|---|---|
| 1 | 309 | 28,09 | T. Chiang | An empirical analysis of herd behavior in global stock markets | 2010 | Journal of Banking and Finance |
| 2 | 238 | 14 | S. Hwang | Market stress and herding | 2004 | Journal of Empirical Finance |
| 3 | 193 | 8,77 | G. Kaminsky | What triggers market jitters?: A chronicle of the Asian crisis | 1999 | Journal of International Money and Finance |
| 4 | 167 | 6,68 | G. Calvo | Mexico's balance-of-payments crisis: A chronicle of a death foretold | 1996 | Journal of International Economics |
| 5 | 165 | 8,68 | W. Kim | Foreign portfolio investors before and during a crisis | 2002 | Journal of International Economics |
| 6 | 120 | 12 | F.Economou | Cross-country effects in herding behaviour: Evidence from four south European markets | 2011 | Journal of International Financial Markets, Institutions and Money |
| 7 | 74 | 12,33 | E.C. Galariotis | Herding on fundamental information: A comparative study | 2015 | Journal of Banking and Finance |
| 8 | 72 | 7,20 | D. Kenourgios | Equity market integration in emerging Balkan markets | 2011 | Research in International Business and Finance |
| 9 | 63 | 7,88 | N. Philippas | Herding behavior in REITs: Novel tests and | 2013 | International Review of |

| | | | | the role of financial crisis | | Financial Analysis |
|---|---|---|---|---|---|---|
| 10 | 61 | 8,71 | A. Mobarek | A cross-country analysis of herd behavior in Europe | 2014 | Journal of International Financial Markets, Institutions and Money |
| 11 | 61 | 3,21 | D. Sornette | A nonlinear super-exponential rational model of speculative financial bubbles | 2002 | International Journal of Modern Physics C |
| 12 | 52 | 3,06 | M. Bowe | Investor herding during financial crisis: A clinical study of the Jakarta Stock Exchange | 2004 | Pacific Basin Finance Journal |
| 13 | 52 | 4,33 | S.Khan | Contagion in the stock markets: The Asian financial crisis revisited | 2009 | Journal of Asian Economics |
| 14 | 50 | 3,85 | M. Cipriani | Herd behavior and contagion in financial markets | 2008 | B.E. Journal of Theoretical Economics |

Source: data processed with PoP, 2020

Table 6 Details of data per cluster

| Cluster | The keywords that appear the most | Keywords | Article |
|---|---|---|---|
| Cluster 1` | Herd Behavior (9) | Asymmetry behavior of financial behavior, financial contagion, financial crisis, frontier market, Pakistani stock market, herd behavior | Herd's behavior (Cipriani, 2008; Khan, 2009; Bharti, 2010 Mobarek, 2014; Huang, 2015; Which, 2015; Galariotis, 2016; Lee, 2017, Litimi, 2017; Vo, 2017; Balcilar, 2017; Shanta,2019; Vo, 2019; Abdeldayem, 2020; Nakagawa, 2020) |
| Cluster 2 | *Herding (35), Liquidity (3), Quantum regression (5),* | Asymmetry, frontier market, trading volume, volatility, vultures, liquidity, quantile regression | *Herding (Kim, 2002; Hwang, 2004; Khamfula, 2007; An, 2010; Chen, 2012; Kremer, 2013; Philippas, 2013; Messis, 2014; Bohl, 2014; Demirer, 2014; Teng, 2014; Bui, 2015; Babalos, 2015; Huang, 2015; Lulin, 2016; Bekiros, 2017; Stavroyiannis, 2017; Arjoon, 2017; Guney, 2017; Chen, 2018; Benkraiem, 2019; Bahadar, 2019; Chen, 2019; Ganesh, 2019; Stavroyiannis, 2019; Syriopoulas, 2019; Vo, 2019;* |

| | | | Nakagawa, 2020; Batmunkh, 2020; Shrotyria, 2020; Tsuchiya, 2020) |
|---|---|---|---|
| | | | *Liquidity (Galariotis, 2016; Bekiros, 2017; Shrotryia, 2020; Bharti, 2020; Duygun, 2021)* |
| | | | *Quantile Regression (Bekiros, 2017; Shantha, 2019; Shrotyria, 2020; Bharti, 2020; Duygun, 2021)* |
| Cluster 3 | *Cross sectional disperse (4), Global Financial Crisis (5), Herding Behavior(11)* | Covid-19, the oil market, market efficiency. Cross sectional disperse, global financial crisis, herding behavior | Cross Sectional disperse (Oarda, 2013; Dos Santos, 2017; Lee, 2017; Shanta,2019) Global financial crisis ( Bohl, 2014; Bharti, 2020, Chang, 2020; Gabbori, 2021) Herding behavior ( Calvo, 1996; Kaminsky, 1999; Chiang, 2010; Klein, 2013; Oarda, 2013; Kulvanich, 2013; Philippas, 2013; Bohl, 2014; Demirer, 2014; Teng, 2014; Javaira, 2015; Lee, 2015; Zheng, 2015; Lugo, 2015; Sharma, 2015; Huang, 2017; Shah, 2017; BenMabrouk, 2018; Yousaf, 2018; Lu, 2018; Bahadar, 2019; Spelta, 2019; Stavroyiannis, 2019; Abeldayem, 2020; Batmunkh, 2020; Tsuchiya, 2020; Yossef, 2020; Duygun, 2021) |
| Cluster 4 | Contagion (3), Financial markets (3), Herding behavior(21), Institutional investors (3). | Commodity markets. Contagion, Financial Markets, Herding Behavior, Institutional Investors. | Transmission (Kaminsky, 1999; Cipriani, 2008; Khan, 2009; Messis, 2014; Teng, 2014; Wahyudi, 2018; Troug, 2020) Financial Markets (Kaminsky, 1999; Cipriani, 2008; Khan, 2009; Economou, 2011; Oarda, 2013; Liu, 2014, Spelta, 2019) Herding behavior ((Calvo, 1996; Kaminsky, 1999; Chiang, 2010; Klein, 2013; Oarda, 2013; Kulvanich, 2013; Philippas, 2013; Bohl, 2014; Demirer, 2014; Teng, 2014; Javaira, 2015; Lee, 2015; Zheng, 2015; Lugo, 2015; Sharma, 2015; Huang, 2017; Shah, 2017; BenMabrouk, 2018; Yousaf, 2018; Lu, 2018; Bahadar, 2019; Spelta, 2019; Stavroyiannis, 2019; Abeldayem, 2020; Batmunkh, 2020; Tsuchiya, 2020; Yossef, 2020; Duygun, 2021) Institutional investors (Chen, 2012; Bohl, 2014; Zheng, 2015; Cai, 2019) |

| Cluster 5 | Crisis (4), European sovereign debt (3), Feedback trading (3), | Disperse return, stock market. Crisis, European sovereign debt, feedback trading. | Crisis (Bohl, 2014; Bohl, 2014; Mesly, 2018; Stavroyiannis, 2019) European sovereign debt (Dos Santos, 2017; Tsuchiya, 2020). Feedback Trading (Kim, 2002; Bowe, 2004; Hsieh, 2011; Ramli, 2016) |
|---|---|---|---|
| Cluster 6 | Emerging stock markets (4), Basic information (3), | *Contrarian, momentum, china, emerging stock market, basic information* | Emerging stock markets ( Chen, 2012; Demirer, 2014; Bui, 2015; Which, 2015; Shah, 2017; Huang, 2017; Indars, 2019; Vo, 2019) Basic Information (Galariotis, 2015; Indars, 2019; Duygun, 2021) |
| Cluster 7 | Financial crisis (21), | Disperse returns on equity, irrational herding, financial crises | Financial Crisis (Khan, 2009; Chiang, 2010; Hsieh, 2011; Chen, 2012; Klein, 2013; Oarda, 2013; Philippas, 2013; Mobarek, 2014; Bowe, 2014; Bohl, 2014; Teng, 2014; Economou, 2011; Galariotis, 2015; Huang, 2015; Sharma, 2015; Which, 2015; Huang, 2017; Stavroyiannis, 2017; Ben Mabrouk, 2018; Kabir, 2018; Yousaf, 2018; Bahadar, 2019; Vo, 2019; Bharti, 2020; Chang, 2020; Jirasakuldech, 2020; Nakagawa, 2020; Tsuchiya, 2020; Fang, 2021; Duygun, 2021; Gabbori, 2021) |

Source: data processed with VOSviewer, 2020

**Conclusion**

The study reviewed a set of 98 articles with themes related to herding behavior in times of crisis. Articles are collected from the database by using PoP software. 98 of these articles are extracted from 165 larger original articles. In this study, researchers concluded that journals with scopus index Q1 /Q2 had a more significant impact in the field of herding behavior in times of crisis than the initial results and processed results. The gaps in this study indicate a direction for future research agendas. As well as summarizing and supporting the important findings of the review. Overall, studying the concept of herding behavior is on an increasing trend, with particular attention requiring to involve more other research collaborations.

The limitations of this research are based on a limited set of keywords and have limited potential by existing databases in PoP software.

# References

Abdeldayem, M. M., & Al Dulaimi, S. H. (2020). Investors Herding Behavior and Pandemic-Risk Related in the Gcc Stock Markets. *Academy of Accounting and Financial Studies Journal*, *24*(1), 1–14. https://www.scopus.com/inward/record.uri?partnerID=HzOxMe3b&scp=85099434342&origin=inward

Abdeldayem, M. M., & Al Dulaimi, S. H. (2020). Investors' herd behavior related to the pandemic-risk reflected on the GCC stock markets. *Zbornik Radova Ekonomskog Fakultet Au Rijeci*, *38*(2), 563–584. https://doi.org/10.18045/zbefri.2020.2.563

An, X., Cordell, L., & Nichols, J. B. (2020). Reputation, Information, and Herding in Credit Ratings: Evidence from CMBS. *Journal of Real Estate Finance and Economics*, *61*(3), 476–504. https://doi.org/10.1007/s11146-019-09701-3

Arjoon, V., & Bhatnagar, C. S. (2017). Dynamic herding analysis in a frontier market. *Research in International Business and Finance*, *42*, 496–508. https://doi.org/10.1016/j.ribaf.2017.01.006

Babalos, V., & Stavroyiannis, S. (2015). Herding, anti-herding behaviour in metal commodities futures: a novel portfolio-based approach. *Applied Economics*, *47*(46), 4952–4966. https://doi.org/10.1080/00036846.2015.1039702

Bahadar, S., Mahmood, H., & Zaman, R. (2019). The Herds of Bulls and Bears in Leveraged ETF Market. *Journal of Behavioral Finance*, *20*(4), 408–423. https://doi.org/10.1080/15427560.2019.1553177

Balcılar, M., Demirer, R., & Ulussever, T. (2017). Does speculation in the oil market drive investor herding in emerging stock markets? *Energy Economics*, *65*, 50–63. https://doi.org/10.1016/j.eneco.2017.04.031

Batmunkh, M. U., Choijil, E., Vieito, J. P., Espinosa-Méndez, C., & Wong, W. K. (2020). Does herding behavior exist in the Mongolian stock market? *Pacific Basin Finance Journal*, *62*, 101352. https://doi.org/10.1016/j.pacfin.2020.101352

Bekiros, S., Jlassi, M., Lucey, B., Naoui, K., & Uddin, G. S. (2017). Herding behavior, market sentiment and volatility: Will the bubble resume? *North American Journal of Economics and Finance*, *42*, 107–131. https://doi.org/10.1016/j.najef.2017.07.005

Benkraiem, R., Bouattour, M., Galariotis, E., & Miloudi, A. (2019). Do investors in SMEs herd? Evidence from French and UK equity markets. *Small Business Economics*. https://doi.org/10.1007/s11187-019-00284-0

BenMabrouk, H. (2018). Cross-herding behavior between the stock market and the crude oil market during financial distress: Evidence from the New York stock exchange. *Managerial Finance*, *44*(4), 439–458. https://doi.org/10.1108/MF-09-2017-0363

Bernales, A., Verousis, T., & Voukelatos, N. (2020). Do investors follow the herd in option markets? *Journal of Banking and Finance*, *119*. https://doi.org/10.1016/j.jbankfin.2016.02.002

Bharti, B., & Kumar, A. (2020). Herding in fast moving consumer group sector: Equity market asymmetry and crisis. *Journal of Asian Finance, Economics and Business*, *7*(9), 39–49. https://doi.org/10.13106/JAFEB.2020.VOL7.NO9.039

Bohl, M. T., Klein, A. C., & Siklos, P. L. (2014). Short-selling bans and institutional investors' herding behaviour: EVIDENCE from the global financial crisis. *International Review of Financial Analysis*, *33*, 262–269. https://doi.org/10.1016/j.irfa.2014.03.004

Bowe, M., & Domuta, D. (2004). Investor herding during financial crisis: A clinical study of the Jakarta Stock Exchange. *Pacific Basin Finance Journal*, *12*(4), 387–418. https://doi.org/10.1016/j.pacfin.2003.09.003

Bui, N. D., Nguyen, L. T. B., & Nguyen, N. T. T. (2015). Herd behaviour in Southeast Asian stock markets - An empirical investigation. *Acta Oeconomica*, *65*(3), 413–429. https://doi.org/10.1556/032.65.2015.3.4

Cai, F., Han, S., Li, D., & Li, Y. (2019). Institutional herding and its price impact: Evidence from the corporate bond market. *Journal of Financial Economics*, *131*(1), 139–167. https://doi.org/10.1016/j.jfineco.2018.07.012

Calvo, G. A., & Mendoza, E. G. (1996). Mexico's balance-of-payments crisis: A chronicle of a death foretold. *Journal of International Economics*, *41*(3–4), 235–264. https://doi.org/10.1016/S0022-1996(96)01436-5

Chang, C. L., McAleer, M., & Wang, Y. A. (2020). Herding behaviour in energy stock markets during the Global Financial Crisis, SARS, and ongoing COVID-19*. *Renewable and Sustainable Energy Reviews*, *134*. https://doi.org/10.1016/j.rser.2020.110349

Chen, Q., Hua, X., & Jiang, Y. (2018). Contrarian strategy and herding behaviour in the Chinese stock market. *European Journal of Finance*, *24*(16), 1552–1568. https://doi.org/10.1080/1351847X.2015.1071715

Chen, Y. F., Yang, S. Y., & Lin, F. L. (2012). Foreign institutional industrial herding in Taiwan stock market. *Managerial Finance*, *38*(3), 325–340. https://doi.org/10.1108/03074351211201442

Chen, Z., Matousek, R., Stewart, C., & Webb, R. (2019). Do rating agencies exhibit herding behaviour? Evidence from sovereign ratings. *International Review of Financial Analysis*, *64*, 57–70. https://doi.org/10.1016/j.irfa.2019.04.011

Chiang, T. C., & Zheng, D. (2010). An empirical analysis of herd behavior in global stock markets. *Journal of Banking and Finance*, *34*(8), 1911–1921. https://doi.org/10.1016/j.jbankfin.2009.12.014

Choudhary, K., Singh, P., & Soni, A. (2019). Relationship Between FIIs' Herding and Returns in the Indian Equity Market: Further Empirical Evidence. *Global Business Review*. https://doi.org/10.1177/0972150919845223

Cipriani, M., & Guarino, A. (2008). Herd behavior and contagion in financial markets. *B.E. Journal of Theoretical Economics*, *8*(1). https://doi.org/10.2202/1935-1704.1390

Clements, A., Hurn, S., & Shi, S. (2017). An empirical investigation of herding in the U.S. stock market. *Economic Modelling*, *67*, 184–192. https://doi.org/10.1016/j.econmod.2016.12.015

Dang, H. V., & Lin, M. (2016). Herd mentality in the stock market: On the role of idiosyncratic participants with heterogeneous information. *International Review of Financial Analysis*, *48*, 247–260. https://doi.org/10.1016/j.irfa.2016.10.005

Demirer, R., Kutan, A. M., & Zhang, H. (2014). Do ADR investors herdα: Evidence from advanced and emerging markets. *International Review of Economics and Finance*, *30*, 138–148. https://doi.org/10.1016/j.iref.2013.10.006

Dos Santos, L. G. G., & Lagoa, S. (2017). Herding behaviour in a peripheral European stock market: The impact of the subprime and the European sovereign debt crises. *International Journal of Banking, Accounting and Finance*, *8*(2), 174–203. https://doi.org/10.1504/IJBAAF.2017.087074

Duarte Duarte, J. B., Garcés Carreño, L. D., & Sierra Suárez, K. J. (2016). Herding effect in economic sectors of the Latin American stock markets: A pre and post-subprime crisis vision. *Contaduría y Administración*, *61*(2), 298–323. https://doi.org/10.1016/j.cya.2015.12.002

Duygun, M., Tunaru, R., & Vioto, D. (2021). Herding by corporates in the US and the Eurozone through different market conditions. *Journal of International Money and Finance*, *110*, 102311. https://doi.org/10.1016/j.jimonfin.2020.102311

Economou, F., Gavriilidis, K., Kallinterakis, V., & Yordanov, N. (2014). Do fund managers herd in frontier markets - And why? *International Review of Financial Analysis*, *40*, 76–87. https://doi.org/10.1016/j.irfa.2015.03.017

Economou, F., Katsikas, E., & Vickers, G. (2016). Testing for herding in the Athens Stock Exchange during the crisis period. *Finance Research Letters*, *18*, 334–341. https://doi.org/10.1016/j.frl.2016.05.011

Economou, F., Kostakis, A., & Philippas, N. (2011). Cross-country effects in herding behaviour: Evidence from four south European markets. *Journal of International Financial Markets, Institutions and Money*, *21*(3), 443–460. https://doi.org/10.1016/j.intfin.2011.01.005

Espinosa-Méndez, C., & Arias, J. (2020). Herding Behaviour in Asutralian stock market: Evidence on COVID-19 effect. *Applied Economics Letters*. https://doi.org/10.1080/13504851.2020.1854659

Fahimnia, B., Sarkis, J. dan Davarzani, H. (2015) 'Green supply chain management: a review andanalisis bibliometri ', *Jurnal Internasional Ekonomi Produksi* , Vol. 162, hlm. 101–114, Elsevier

Fang, H., Lu, Y.-C., Shieh, J. C. P., & Lee, Y.-H. (2021). The existence and motivations of irrational loan herding and its impact on bank performance when considering different market periods. *International Review of Economics & Finance*, *73*, 420–443. https://doi.org/10.1016/j.iref.2021.01.015

Fink, A. (2005) *Melakukan Riset Literatur Review. From the Internet to Paper* , edisi ke-4, SAGE Publications, Los Angeles.

Frenkel, M., Mauch, M., & Rülke, J. C. (2020). Do forecasters of major exchange rates herd? *Economic Modelling*, *84*, 214–221. https://doi.org/10.1016/j.econmod.2019.04.011

Gabbori, D., Awartani, B., Maghyereh, A., & Virk, N. (2020). OPEC meetings, oil market volatility and herding behaviour in the Saudi Arabia stock market. *International Journal of Finance and Economics*, *26*(1), 870–888. https://doi.org/10.1002/ijfe.1825

Galariotis, E. C., Krokida, S. I., & Spyrou, S. I. (2016). Herd behavior and equity market liquidity: Evidence from major markets. *International Review of Financial Analysis*, *48*, 140–149. https://doi.org/10.1016/j.irfa.2016.09.013

Galariotis, E. C., Krokida, S. I., & Spyrou, S. I. (2016). Bond market investor herding: Evidence from the European financial crisis. *International Review of Financial Analysis*, *48*, 367–375. https://doi.org/10.1016/j.irfa.2015.01.001

Galariotis, E. C., Rong, W., & Spyrou, S. I. (2015). Herding on fundamental information: A comparative study. *Journal of Banking and Finance*, *50*, 589–598. https://doi.org/10.1016/j.jbankfin.2014.03.014

Ganesh, R., Naresh, G., & Thiyagarajan, S. (2019). Institutional herding in sensitivity index. *Finance India*, *33*(4), 967–980. https://www.scopus.com/inward/record.uri?partnerID=HzOxMe3b&scp=85079870532&origin=inward

Garza-Reyes, JA (2015) 'Lean and green - a sistematis review of state of the art literature',*Jurnal Produksi Bersih* , Vol. 102, hlm. 18–29 [online] https://doi.org/10.1016/ j.jclepro.2015.04.064 (diakses 7 Maret 2017)

Guney, Y., Kallinterakis, V., & Komba, G. (2017). Herding in frontier markets: Evidence from African stock exchanges. *Journal of International Financial Markets, Institutions and Money*, *47*, 152–175. https://doi.org/10.1016/j.intfin.2016.11.001

Heersmink, R., Hoven, J., Van Den, E., Van Den, NJ dan Berg, J. (2010) *Pemetaan Bibliometrik Etika Komputer dan Informasi* , Seri Kertas Kerja CWTS

Hsieh, M. F., Yang, T. Y., Yang, Y. T., & Lee, J. S. (2011). Evidence of herding and positive feedback trading for mutual funds in emerging Asian countries. *Quantitative Finance*, *11*(3), 423–435. https://doi.org/10.1080/14697688.2010.506882

Huang, T. C., Lin, B. H., & Yang, T. H. (2015). Herd behavior and idiosyncratic volatility. *Journal of Business Research*, *68*(4), 763–770. https://doi.org/10.1016/j.jbusres.2014.11.025

Huang, T. C., & Wang, K. Y. (2017). Investors' Fear and Herding Behavior: Evidence from the Taiwan Stock Market. *Emerging Markets Finance and Trade*, *53*(10), 2259–2278. https://doi.org/10.1080/1540496X.2016.1258357

Humayun Kabir, M. (2018). Did Investors Herd during the Financial Crisis? Evidence from the US Financial Industry. *International Review of Finance*, *18*(1), 59–90. https://doi.org/10.1111/irfi.12140

Hwang, S., & Salmon, M. (2004). Market stress and herding. *Journal of Empirical Finance*, *11*(4), 585–616. https://doi.org/10.1016/j.jempfin.2004.04.003

Indārs, E. R., Savin, A., & Lublóy, Á. (2019). Herding behaviour in an emerging market: Evidence from the Moscow Exchange. *Emerging Markets Review*, *38*, 468–487. https://doi.org/10.1016/j.ememar.2018.12.002

Javaira, Z., & Hassan, A. (2015). An examination of herding behavior in Pakistani stock market. In *International Journal of Emerging Markets* (Vol. 10, Issue 3, pp. 474–490). thesis.cust.edu.pk. https://doi.org/10.1108/IJoEM-07-2011-0064

Jirasakuldech, B., & Emekter, R. (2020). Empirical Analysis of Investors' Herding Behaviours during the Market Structural Changes and Crisis Events: Evidence from Thailand. *Global Economic Review*. https://doi.org/10.1080/1226508X.2020.1821746

Kaminsky, G. L., & Schmukler, S. L. (1999). What triggers market jitters?: A chronicle of the Asian crisis. In *Journal of International Money and Finance* (Vol. 18, Issue 4). elibrary.worldbank.org. https://doi.org/10.1016/S0261-5606(99)00015-7

Kashif, M., Palwishah, R., Ahmed, R. R., Vveinhardt, J., & Streimikiene, D. (2020). Do investors herd? An examination of Pakistan stock exchange. *International Journal of Finance and Economics*, *n/a*(n/a). https://doi.org/10.1002/ijfe.1895

Kenourgios, D., & Samitas, A. (2011). Equity market integration in emerging Balkan markets. *Research in International Business and Finance*, *25*(3), 296–307. https://doi.org/10.1016/j.ribaf.2011.02.004

Khamfula, Y., Mlachila, M., & Chirwa, E. (2007). Donor herding and domestic debt crisis. *Applied Economics Letters*, *14*(4), 299–302. https://doi.org/10.1080/13504850500447356

Khan, S., & Park, K. W. (Ken). (2009). Contagion in the stock markets: The Asian financial crisis revisited. *Journal of Asian Economics*, *20*(5), 561–569. https://doi.org/10.1016/j.asieco.2009.07.001

Kim, W., & Wei, S. J. (2002). Foreign portfolio investors before and during a crisis. *Journal of International Economics*, *56*(1), 77–96. https://doi.org/10.1016/S0022-1996(01)00109-X

Klein, A. C. (2013). Time-variations in herding behavior: Evidence from a Markov switching SUR model. *Journal of International Financial Markets, Institutions and Money*, *26*(1), 291–304. https://doi.org/10.1016/j.intfin.2013.06.006

Kremer, S., & Nautz, D. (2013). Short-term herding of institutional traders: New evidence from the German stock market. *European Financial Management*, *19*(4), 730–746. https://doi.org/10.1111/j.1468-036X.2011.00607.x

Kulvanich, J., & Boonvorachote, T. (2013). Herding behavior analysis in the stock exchange of Thailand. *Kasetsart Journal - Social Sciences*, *34*(1), 43–59. https://so04.tci-thaijo.org/index.php/kjss/article/view/246870

Lee, K. (2017). Herd behavior of the overall market: Evidence based on the cross-sectional comovement of returns. *North American Journal of Economics and Finance*, *42*, 266–284. https://doi.org/10.1016/j.najef.2017.07.006

Lee, S., & Lee, K. (2015). Heterogeneous expectations leading to bubbles and crashes in asset markets: Tipping point, herding behavior and group effect in an agent-based model. *Journal of Open Innovation: Technology, Market, and Complexity*, *1*(1). https://doi.org/10.1186/s40852-015-0013-9

Litimi, H. (2017). Herd behavior in the French stock market. *Review of Accounting and Finance*, *16*(4), 497–515. https://doi.org/10.1108/RAF-11-2016-0188

Liu, X. D., Liu, C., Liu, S. C., & Lu, J. J. (2014). Does herd behavior increase stock price volatility? *Xitong Gongcheng Lilun Yu Shijian/System Engineering Theory and Practice*, *34*(6), 1361–1368. https://www.scopus.com/inward/record.uri?partnerID=HzOxMe3b&scp=84904889585&origin=inward

Lu, S., Zhao, J., Wang, H., & Ren, R. (2018). Herding boosts too-connected-to-fail risk in stock market of China. *Physica A: Statistical Mechanics and Its Applications*, *505*, 945–964. https://doi.org/10.1016/j.physa.2018.04.020

Lugo, S., Croce, A., & Faff, R. (2015). Herding behavior and rating convergence among credit rating agencies: Evidence from the subprime crisis. *Review of Finance*, *19*(4), 1703–1731. https://doi.org/10.1093/rof/rfu028

Lulin, Z., Antwi, H. A., Wang, W., Yiranbon, E., Marfo, E. O., & Acheampong, P. (2016). The effect of herd formation among healthcare investors on health sector growth in China. *International Journal for Equity in Health*, *15*(1). https://doi.org/10.1186/s12939-016-0393-x

Luo, Z., & Schinckus, C. (2015). The influence of the US market on herding behaviour in China. *Applied Economics Letters*, *22*(13), 1055–1058. https://doi.org/10.1080/13504851.2014.997920

Mesly, O., & Racicot, F. É. (2018). Heteroscedasticity of deviations in market bubble moments–how the goods and bads lead to the ugly. *Applied Economics*, *50*(32), 3441–3463. https://doi.org/10.1080/00036846.2017.1420902

Messis, P., & Zapranis, A. (2014). Herding towards higher moment CAPM, contagion of herding and macroeconomic shocks: Evidence from five major developed markets. *Journal of Behavioral and Experimental Finance*, *4*, 1–13. https://doi.org/10.1016/j.jbef.2014.09.002

Mobarek, A., Mollah, S., & Keasey, K. (2014). A cross-country analysis of herd behavior in Europe. *Journal of International Financial Markets, Institutions and Money*, *32*(1), 107–127. https://doi.org/10.1016/j.intfin.2014.05.008

Nakagawa, R. (2020). Bank herding in loan markets: Evidence from geographical data in Japan. *International Review of Finance*. https://doi.org/10.1111/irfi.12341

Ouarda, M., El Bouri, A., & Bernard, O. (2013). Herding behavior under markets condition: Empirical evidence on the European financial markets. *International Journal of Economics and Financial Issues*, *3*(1), 214–228. https://www.scopus.com/inward/record.uri?partnerID=HzOxMe3b&scp=84979815376&origin=inward

Persaud, A. (2000). Sending the herd off the cliff edge: The disturbing interaction between herding and market-sensitive risk management practices. *Journal of Risk Finance*, *2*(1), 59–65. https://doi.org/10.1108/eb022947

Philippas, N., Economou, F., Babalos, V., & Kostakis, A. (2013). Herding behavior in REITs: Novel tests and the role of financial crisis. *International Review of Financial Analysis*, *29*, 166–174. https://doi.org/10.1016/j.irfa.2013.01.004

Ramli, I., Agoes, S., & Setyawan, I. R. (2016). Information asymmetry and the role of foreign investors in daily transactions during the crisis; A study of herding in the indonesian stock exchange. *Journal of Applied Business Research*, *32*(1), 269–288. https://doi.org/10.19030/jabr.v32i1.9537

Shah, M. U. D., Shah, A., & Khan, S. U. (2017). Herding behavior in the Pakistan stock exchange: Some new insights. *Research in International Business and Finance*, *42*, 865–873. https://doi.org/10.1016/j.ribaf.2017.07.022

Shantha, K. V. A. (2019). The evolution of herd behavior: Will herding disappear over time? *Studies in Economics and Finance*, *36*(3), 637–661. https://doi.org/10.1108/SEF-06-2018-0175

Sharma, S. S., Narayan, P., & Thuraisamy, K. (2015). Time-varying herding behavior, global financial crisis, and the Chinese stock market. *Review of Pacific Basin Financial Markets and Policies*, *18*(2). https://doi.org/10.1142/S0219091515500095

Shrotryia, V. K., & Kalra, H. (2020). Herding and BRICS markets: a study of distribution tails. *Review of Behavioral Finance*, *ahead-of-p*(ahead-of-print). https://doi.org/10.1108/RBF-04-2020-0086

Sornette, D., & Andersen, J. V. (2002). A nonlinear super-exponential rational model of speculative financial bubbles. *International Journal of Modern Physics C*, *13*(2), 171–187. https://doi.org/10.1142/S0129183102003085

Spelta, A., Flori, A., Pecora, N., & Pammolli, F. (2019). Financial crises: Uncovering self-organized patterns and predicting stock markets instability. *Journal of Business Research*. https://doi.org/10.1016/j.jbusres.2019.10.043

Stavroyiannis, S., & Babalos, V. (2017). Herding, Faith-Based Investments and the Global Financial Crisis: Empirical Evidence From Static and Dynamic Models. *Journal of Behavioral Finance*, *18*(4), 478–489. https://doi.org/10.1080/15427560.2017.1365366

Stavroyiannis, S., & Babalos, V. (2019). Time-varying herding behavior within the Eurozone stock markets during crisis periods: Novel evidence from a TVP model. *Review of Behavioral Finance*, *12*(2), 83–96. https://doi.org/10.1108/RBF-07-2018-0069

Syriopoulos, T., & Bakos, G. (2019). Investor herding behaviour in globally listed shipping stocks. *Maritime Policy and Management*, *46*(5), 545–564. https://doi.org/10.1080/03088839.2019.1597288

Teng, Y. P., & Liu, Y. A. (2014). The study of herding behavior among the Greater China Stock Markets during Financial Crisis. *Journal of Interdisciplinary Mathematics*, *17*(2), 163–197. https://doi.org/10.1080/09720502.2013.878817

Tran, V. T., Nguyen, H., & Lin, C. T. (2017). Herding behaviour in the Australian loan market and its impact on bank loan quality. *Accounting and Finance*, *57*(4), 1149–1176. https://doi.org/10.1111/acfi.12183

Tranfield, D., Denyer, D. dan Smart, P. (2003) 'Menuju metodologi untuk mengembangkan bukti-pengetahuan manajemen yang diinformasikan melalui tinjauan sistematis * ', *British Journal of Manajemen* , Vol. 14, No. 1–3, hlm. 207–222 [online] https://doi.org/10.1111/1467-8551.00375

Troug, H., & Murray, M. (2020). Crisis determination and financial contagion: an analysis of the Hong Kong and Tokyo stock markets using an MSBVAR approach. *Journal of Economic Studies*. https://doi.org/10.1108/JES-03-2020-0095

Tsuchiya, Y. (2020). Crises, market shocks, and herding behavior in stock price forecasts. *Empirical Economics*. https://doi.org/10.1007/s00181-020-01894-4

Vo, X. V., & Phan, D. B. A. (2019). Herding and equity market liquidity in emerging market. Evidence from Vietnam. *Journal of Behavioral and Experimental Finance*, *24*, 100189. https://doi.org/10.1016/j.jbef.2019.02.002

Vo, X. V., & Phan, D. B. A. (2017). Further evidence on the herd behavior in Vietnam stock market. *Journal of Behavioral and Experimental Finance*, *13*, 33–41. https://doi.org/10.1016/j.jbef.2017.02.003

Vo, X. V., & Phan, D. B. A. (2019). Herd behavior and idiosyncratic volatility in a frontier market. *Pacific Basin Finance Journal*, *53*, 321–330. https://doi.org/10.1016/j.pacfin.2018.10.005

Wahyudi, S., Laksana, R. D., Najmudin, N., & Rachmawati, R. (2018). Assessing the contagion effect on herding behaviour under segmented and integrated stock markets circumstances in the USA, China, and ASEAN-5. *Economic Annals-XXI*, *169*(1–2), 15–20. https://doi.org/10.21003/ea.V169-03

Yang, W. R., & Chen, Y. L. (2015). The response of dynamic herd behavior to domestic and u.s. market factors: Evidence from the greater China stock markets. *Emerging Markets Finance and Trade*, *51*, S18–S41. https://doi.org/10.1080/1540496X.2014.998884

Yousaf, I., Ali, S., & Shah, S. Z. A. (2018). Herding behavior in Ramadan and financial crises: the case of the Pakistani stock market. In *Financial Innovation* (Vol. 4, Issue 1). Springer. https://doi.org/10.1186/s40854-018-0098-9

Youssef, M. (2020). Do Oil Prices and Financial Indicators Drive the Herding Behavior in Commodity Markets? *Journal of Behavioral Finance*. https://doi.org/10.1080/15427560.2020.1841193

Youssef, M., & Mokni, K. (2020). Asymmetric effect of oil prices on herding in commodity markets. *Managerial Finance*. https://doi.org/10.1108/MF-01-2020-0028

Zheng, D., Li, H., & Zhu, X. (2015). Herding behavior in institutional investors: Evidence from China's stock market. *Journal of Multinational Financial Management*, *32–33*, 59–76. https://doi.org/10.1016/j.mulfin.2015.09.001